\documentclass[%
 reprint,
 amsmath,amssymb,
 aps,
pre,
]{revtex4-1}
\usepackage{braket}  
\usepackage{graphicx}
\usepackage{dcolumn}
\usepackage{bm}
\usepackage{enumerate}

\begin{document}

\preprint{APS/123-QED}

\title{Dual application of Chebyshev polynomial for efficiently \\computing thousands of central eigenvalues in many-spin systems}

\author{Haoyu Guan}
\affiliation{%
	School of Physics and Technology, Wuhan University, Wuhan, Hubei 430072, China
}%
\author{Wenxian Zhang}%
\email{wxzhang@whu.edu.cn}
\affiliation{%
	School of Physics and Technology, Wuhan University, Wuhan, Hubei 430072, China
}%

\date{\today}

\begin{abstract}
Computation of a large group of interior eigenvalues at the middle spectrum is an important problem for quantum many-body systems, where the level statistics provides characteristic signatures of quantum chaos. We propose an efficient numerical method, dual application of Chebyshev polynomial (DACP), to find thousands of central eigenvalues, which are exponentially close to each other in terms of the system size. To cope with the near-degenerate problem, we employ twice the Chebyshev polynomial to construct an exponential semicircle filter as a preconditioning step and generate a large set of  proper states as the basis of the desired subspace. Numerical experiments on Ising spin chain and spin glass shards confirm the correctness and efficiency of the DACP method. As numerical results demonstrate, DACP is $30$ times faster than the state-of-the-art shift-invert method for the Ising spin chain while $8$ times faster for the spin glass shards. In contrast to shift-invert method, the computation time of DACP is  only weakly influenced by the required number of eigenvalues. Moreover, the consumed memory remains a constant for spin-$1/2$ systems consisting of up to $20$ spins.
\end{abstract}

\maketitle
\section{introduction}
Energy level statistics provides an essential characterization of quantum chaos~\cite{QuantumChaos,RevModPhys.53.385}. Integrable systems often imply level clustering and a Poisson distribution of energy level spacings~\cite{RevModPhys.53.385}, while chaotic systems exhibit strong level repulsion and a Wigner-Dyson distribution~\cite{PhysRevLett.52.1}. Other useful statistical tools like the $\delta_3$ statistic~\cite{random} and the power spectrum of the $\delta_n$ statistic also depend on the level distribution~\cite{PhysRevLett.89.244102}. Numerical observation of eigenvalues is important to exactly characterize these level statistics, because direct theoretical derivation is not available. In addition, the need to access individual eigenstates in the middle of the spectrum, which correspond to the ``infinite temperature" limit, is emphasized in studying the phenomenon of many-body localization (MBL)~\cite{BASKO20061126,MBL,scipost}. Numerical simulations remain instrumental in understanding quantitatively many aspects of the MBL problem.
As the many-body problems of interest often involve a huge Hilbert space dimension growing exponentially in terms of system size, it is rather challenging to conduct a full diagonalization of the Hamiltonian or to solve the time-independent Schr\"{o}dinger equation. Seeking to resolve the eigenvalue problem in a small part of the spectrum is thus an unavoidable and desirable substitution.

To obtain the interior eigenstates, a hybrid strategy of matrix spectroscopy is often invoked~\cite{Thomas1980The,PhysRevE.51.3643,PhysRevE.56.4837}. It aims at computing eigenstates in selected regions of the spectrum and couples the ground state solvers with a spectral filter, where the filter is designed to cast the selected interior regions to the edges of the spectrum.
Among these filters, the Green function $(\mathcal{H}-\lambda I)^{-1}$ is an excellent one. After applying the Green function filter, the cluster of eigenvalues near the test energy $\lambda$ is mapped to very large positive and negative values. The level spacings near $\lambda$ are greatly amplified, which thus improves the convergence. The Lanczos method~\cite{Lanczos2018An} has been coupled with such a filter~\cite{Thomas1980The,PhysRevE.51.3643}, and the Chebyshev polynomial expansion of the Green function is implemented~\cite{doi:10.1063/1.470477}. In particular, the shift-invert method~\cite{TSA} essentially utilizes this spectral transformation and has been widely used in quantum many-spin systems~\cite{scipost,PhysRevB.91.081103,PhysRevB.101.104201,PhysRevA.101.063617,PhysRevB.102.064207}. Moreover, it was even proved to be the most efficient one for the MBL problem~\cite{scipost}. However, for large systems this method suffers rapid increases in both computation time and memory consumption, due to the factorization of $(\mathcal{H}-\lambda I)^{-1}$. Other spectral filters have been also proposed, including the Dirac delta function filter $\delta(\mathcal{H}-\lambda I)$~\cite{TRA, dd, PhysRevLett.125.156601, Guan} and the rectangular window function filter~\cite{Pieper2016High}, both of which are expanded by the Chebyshev polynomial. Note that all the methods mentioned above are iterative ones, where the computation time is usually proportional to the required number of eigenvalues, i.e., more eigenvalues means more filtrations and reorthogonalizations~\cite{scipost,dd,PhysRevLett.125.156601}.

In this paper, we propose an exact numerical method, DACP, to calculate thousands of eigenvalues in the middle of the energy band, which is already helpful to reveal the level statistics~\cite{PhysRevB.47.11487,sgs}. For spin systems, such a region usually locates at the peak of the density of states where the energy levels are so close that they are nearly degenerate. It is extremely challenging to distinguish these near-degenerate eigenvalues without factorizing $\mathcal{H}^{-1}$. In the DACP method, we construct an exponential of semicircle (exp-semicircle) filter to quickly damp the unwanted part of the spectrum by employing the Chebyshev polynomial for the first time. The second application of the Chebyshev polynomial is to fast search a set of states to span the specific subspace consisting of all the eigenstates corresponding to the desired eigenvalues. The DACP method transforms the original high dimensional eigenvalue problem to a low dimensional one. For practical problems in many-spin systems, the DACP method is very efficient, due to its full exploration of several excellent properties of the Chebyshev polynomial while other methods usually utilize only one of them. For a large class of many-spin systems, the DACP method exhibits a significant increase in the computation speed, up to a factor of $30$, in comparison with the shift-invert method. The memory saving is more drastic, up to a factor of $100$. Moreover, the DACP method distinguishes itself from those iterative filtering ones, as its convergence time varies only slightly when the required number of eigenvalues is vastly changed.

The paper is organized as follows. The detailed formalism of the DACP method, including the exp-semicircle filtration, Chebyshev evolution, and subspace diagonalization, are described in Sec.~\ref{sec:TD}. Each of the processes relies on a property of the Chebyshev polynomial. In Sec.~\ref{sec:NR} we discuss details of two many-spin systems, the Ising model and the spin glass shards, and present the computational results. We conclude in Sec.~\ref{sec:CON}.

\section{Dual application of Chebyshev polynomial method} \label{sec:TD}

To access central eigenvalues of large spin systems, we are restricted by the matrix-free mode, i.e., the matrix of the Hamiltonian must not be explicitly expressed/stored. Instead, we treat the Hamiltonian $\mathcal{H}$ as a function whose input and output are both states/vectors. Therefore, we shall only operate with the quantum states $\left|\psi\right>, \mathcal{H}\left|\psi\right>, \cdots, \mathcal{H}^k\left|\psi\right>$, where $k$ is a positive integer (not too large), in the original Hilbert space  of dimension up to $5\times 10^5$. In this paper, we set our goal to find $5,000$ eigenvalues in the middle of the spectrum for many-spin systems as an illustration.

The DACP idea is fairly straightforward. We first transform a randomly initialized state into a wave packet in the subspace spanned essentially by at least $5,000$ central eigenstates. This transformation is realized by an exp-semicircle filter implemented by the Chebyshev polynomial. The Chebyshev polynomial is used to perform an exponential decay. With this particular state in hand, we then generate a large amount of linearly independent states, as large as possible, to approximately span the subspace consisting of the required eigenstates. The Chebyshev polynomial is used again to oscillate (complex exponential) the states, producing approximately linearly independent states. Once the generating set for the desired eigenstates is known, one may explicitly calculate $H$, the reduced representation of $\mathcal{H}$ in this subspace. The remaining operations are restricted in the subspace of dimension around $10^4$. Finally, direct diagonalization of $H$, which is of size $10^4\times 10^4$, gives the required eigenvalues. The detailed procedures and discussions are given in the subsections below.

\subsection{Exp-semicircle filter}
 \begin{figure}[b]
	\includegraphics[width=3.25 in]{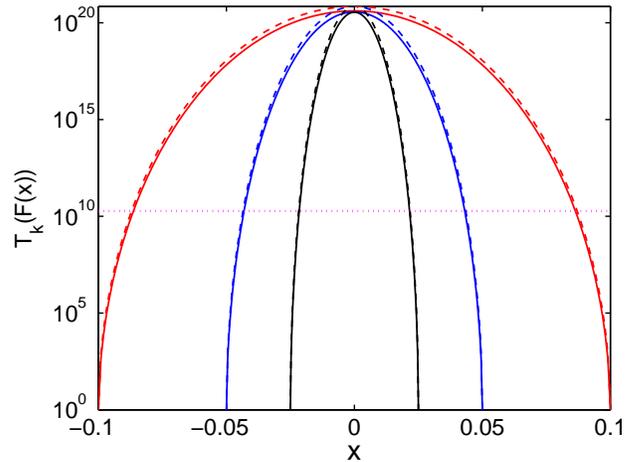}
	\caption{\label{fig:filter} (Color online.) The exp-semicircle filter for $a=0.1$ (red lines), $0.05$ (blue lines), and $0.025$ (black lines), with $ka=24$. We have set $E_{\rm max}=1$ and $E_{\rm min}=-1$. The solid lines are for the Chebyshev polynomial $T_k(F(x))$, where $F(x)=\left[2x^2-(1+a^2)\right]/(1-a^2)$, while the dashed lines are for the approximation $y=\exp(2k\sqrt{a^2-x^2})\simeq T_k(F(x))$. The horizontal pink dotted line denotes the half maximum of the logarithm of filters. Note that $|T_k(F(x))|\le 1$ when $x\notin [-a,a]$, thus the filter graphs are restricted in $[-a,a]$.}
\end{figure}
We utilize the exponential growth of the Chebyshev polynomial outside the interval $[-1,1]$ to efficiently construct an exp-semicircle filter, as shown in Fig.~\ref{fig:filter}. This filter drastically amplifies the components of a desired range of eigenstates for any randomly initialized states, resulting a new state that sharply localized in the central spectrum. We note that the Chebyshev filter explores the same property as well, except that it amplifies the lower end of the spectrum~\cite{cd1,cd2}. A similar idea has been also applied in the quantum algorithm for finding ground states~\cite{PhysRevLett.102.130503}.

For the Hamiltonian $\mathcal H$ with energy bounded in $[E_{\rm min}, E_{\rm max}]$, where $E_{\rm min}$ is the minimum energy ($E_{\rm min}<0$) and $E_{\rm max}$ the maximum energy ($E_{\rm max}>0$), the exp-semicircle filter is designed to amplify the components of the eigenstates corresponding to eigenvalues in the interval $[-a,a]$ and to simultaneously dampen those in the interval $[E_{\rm min},-a]$ and $[a,E_{\rm max}]$, where $a$ is a real positive parameter. Focusing on the spin systems, we have assumed $E_{\rm min}<-a$ and $a<E_{\rm max}$. After the filtration, a new state mainly consists of the eigenstates with eigenvalues belonging to the interval $[-a,a]$ is generated. For simplicity, let us denote the subspace spanned by the eigenstates contained in $[-a,a]$ as $\mathbb{L}$.

We now introduce the specific implementation details. Note that we want to amplify the middle of the spectrum, but the exponential growth of the Chebyshev polynomial exists only in both ends. To satisfy this goal, we first square the Hamiltonian $\mathcal{H}^2$ whose spectrum ranges $[0,E_{\rm max}^2]$ (suppose $E_{\rm min}=-E_{\rm max}$ for simplicity). The central spectrum $[-a,a]$ of $\mathcal{H}$ is transferred to $[0,a^2]$ for $\mathcal{H}^2$, which lies exactly at the lower end. Next is to map the dampening part $[a^2,E_{\rm max}^2]$ into $[-1,1]$ by shift and normalization of $\mathcal{H}^2$. We thus define an  operator
\begin{equation}
\mathcal{F}=\frac{\mathcal{H}^2-E_c}{E_0},
\label{eq:F}
\end{equation}
where $E_c=(E_{\rm max}^{2}+a^2)/2$
and $E_0=(E_{\rm max}^{2}-a^2)/2$. One may easily affirm this map's correctness by replacing $\mathcal{H}^2$ with either $a^2$ or $E_{\rm max}^{2}$, and correspondingly one has $F(x)=(x^2-E_c)/E_0$. Note that $\mathcal{F}$ is simply a polynomial expression of $\mathcal{H}$, so is $T_k(\mathcal{F})$.

We then explore the effect of the filtration using $T_k(\mathcal{F})$. As the eigenvalues inside $[0,a^2]$ of $\mathcal{H}^2$ is mapped into $[-1-2a^2/(E_{\rm max}^2-a^2),-1]$ of $\mathcal{F}$, we have $T_k(\mathcal{F})=(-1)^k \cosh(k\Theta)$ for the lower end of the spectrum, where
\begin{equation}
	\Theta=\cosh^{-1}\left(1+\frac{2(a^2-\mathcal{H}^2)}{E_{\rm max}^2-a^2}\right).
\end{equation}
Let $\left|\psi\right>=\sum_{i}{c_i \left| \phi _i \right>}+\sum_{j}{d_j \left| \chi _j \right>}$ be a random initial state, with $c_i$ and $d_j$ the random coefficients, $\left| \phi _i \right>$ the eigenstates inside $\mathbb{L}$, $\left| \chi _j \right>$ the eigenstates outside $\mathbb{L}$. The filtration by $T_k(\mathcal{F})$ is
\begin{equation}
\label{eq:chevo}
\begin{aligned}
\left| \psi \left( k \right) \right>
&=T_k\left( \mathcal{F} \right) \left| \psi \right> \\
&=\sum_{i}{\left( {\rm e}^{k\theta _i^{\rm in}}+{\rm e}^{-k\theta _i^{\rm in}} \right) \frac{c_i}{2}\left| \phi _i \right>} \\
&\quad+\sum_{j}{ \left( {\rm e}^{ik\theta _j^{\rm out}}+{\rm e}^{-ik\theta_j^{\rm out}} \right) \frac{d_j}{2}\left| \chi _j \right>}\\
& \simeq \frac{1}{2}\sum_{i}{{\rm e}^{k\theta _i^{\rm in}} c_i\left| \phi _i \right>},
\end{aligned}
\end{equation}
where $\theta_i^{\rm in}=\cosh^{-1}(1+2(a^2-E_i^2)/(E_{\rm max}^2-a^2))$, $\theta_j^{\rm out}=\cos^{-1}(2(E_j^2-a^2)/(E_{\rm max}^2-a^2)-1)$, $E_i$ and $E_j$ are eigenvalues corresponding to $\left| \phi _i \right>$ and $\left| \chi _j\right>$, respectively. In writing Eq.~(\ref{eq:chevo}) we have ignored $(-1)^k$, as it does not affect the absolute value of coefficients and is a global phase at the end line. When $a$ is tiny, i.e., $a^2\ll E_{\rm max}^2$, one may further deduce $\theta_i^{\rm in}\simeq 2\sqrt{a^2-E_i^2}/E_{\rm max}$ via Taylor's expansion of $\cosh^{-1}(1+\varepsilon)$, where $\varepsilon$ is a small positive number. We thus obtain the exp-semicircle filter \begin{equation}
\label{eq:filter}
T_k(\mathcal{F})\simeq {\rm e}^{\frac{2k}{E_{\rm max}}\sqrt{a^2-\mathcal{H}^2}}
\end{equation}
that peaked sharply at $E_i=0$ with a large $k$, for eigenstates satisfying $-a\le E_i \le a$. In Fig.~\ref{fig:filter} the shape of Eq.~(\ref{eq:filter}) is presented in dashed lines, which fits pretty well with the exact form.

With the initial conditions $T_0(\mathcal{F})=1$ and $T_1(\mathcal{F})=\mathcal{F}$, the $k$th order Chebyshev polynomial can be efficiently determined using the recurrence relation Eq.~(\ref{eq:rec}). In this paper, we set the cut-off order  $K=18E_{\rm max}/a$. Such a filter exponentially (the fastest rate among all polynomials) amplifies the components of eigenstates inside $\mathbb{L}$~\cite{Rivlin}.  After the normalization, it equivalently dampens those outside $\mathbb{L}$, generating the target state:
\begin{equation}\label{eq:target}
\left| \psi_E\right>\simeq\sum_i{c_i^{\prime}\left| \phi_i\right>},
\end{equation}
where $c_i^{\prime}=\beta {\rm e}^{K\theta_i^{\rm in}}c_i$ and $\beta$ is the normalization constant. Obviously, the state $\left| \psi_E\right>$  localizes (in the energy representation) in the central spectrum. We input $\left| \psi_E\right>$ as the initial state for the next step, Chebyshev evolution.

\subsection{Chebyshev evolution}
After obtaining a state confined in the small subspace $\mathbb{L}$, we then make use of the oscillation property of the Chebyshev polynomial to efficiently generate a set of {\it distinct} states, as many as possible, serving as a complete basis to span $\mathbb{L}$. To achieve this goal, it is necessary to limit $E_i\in [-1,1]$ (corresponding to $x$ in $T_k(x)$), within which the Chebyshev polynomial behaves as a cosine-like function. This region contrasts with the requirement of the exp-semicircle filter, thus one needs a different transformation of $\mathcal{H}$. Below we describe the specific details for the second application of the Chebyshev polynomial.

The original Hamiltonian $\mathcal{H}$ needs to be shifted by $E_c^{\prime}$ and be rescaled by $E_0^{\prime}$, where $E_c^{\prime}=\frac{1}{2}(E_{\rm min}+E_{\rm max})$ and $E_0^{\prime}=\frac{1}{2}(-E_{\rm min}+E_{\rm max})$.  In a similar way to Eq.~(\ref{eq:F}), we define an operator
$	
	\mathcal{G}=(\mathcal{H}-E_c^{\prime})/E_0^{\prime},
$
which is definitely bounded by $-1$ and $1$. Assuming $E_{\rm min}=-E_{\rm max}$ again, which does not affect the conclusion, we obtain
\begin{equation}
	\label{eq:G2}
		\mathcal{G}=\frac{\mathcal{H}}{E_{\rm max}}.
\end{equation}
At the same time, the parameter $a$ is rescaled to $a_r=a/E_{\rm max}$ as well.

Let us explore the Chebyshev evolution, i.e., the evolution governed by the operator $T_k(\mathcal{G})$ as $k$ plays the role of time.
We input the state Eq.~(\ref{eq:target}) generated by the filtration as an initial state for the Chebyshev evolution. Since $\Vert \mathcal{G} \Vert\le1$, from Eq.~(\ref{eq:cheb2}) we have $T_k(\mathcal{G})=\cos(k\Omega)$, where
$
	\Omega=\arccos(\mathcal{G})
$.
In this sense, the Chebyshev evolution gives
 \begin{equation}
 \label{eq:chevo2}
 \begin{aligned}
 \left| \psi_E \left( k \right) \right>
 &=T_k\left( \mathcal{G} \right) \left| \psi_E \right> \\
 &=\frac{1}{2}\sum_{j}\left({\rm e}^{{\rm i}k\omega_j}+{\rm e}^{-{\rm i}k\omega_j}\right)c_j^{\prime} \left| \phi_j \right>,
 \end{aligned}
 \end{equation}
where $\omega_j=\arccos(E_j/E_{\rm max})$. Note that both $\omega_j$ and $k$ are unitless. Certainly, this is not a physical evolution, and $T_k(\mathcal{G})$ is not even a unitary operator. Actually, this evolution essentially represents a superposition of both forward and backward propagation. Each time the polynomial order $k$ increases by $1$, the evolution ``time" is added by $1$ as well. 

With the aid of the Chebyshev evolution, we are able to construct a complete basis that spans the subspace $\mathbb{L}$. In detail, we collect a set of states as follows
 \begin{equation}\label{eq:set}
 	\left\{\hat{I},\sin(\hat{X}),\cdots,\sin(n\hat{X}), \cos(\hat{X}),\cdots,\cos(n\hat{X})\right\} \left| \psi_E \right>,
 \end{equation}
where $\hat{X}=$$\pi\mathcal{G}/{a_r}$ and $n$ is an integer determined by the relation $2n+1\ge d$, with $d$ the dimension of $\mathbb{L}$. In practice, one may need a relation $2n+1=1.5d$ to ensure the completeness of Eq.~(\ref{eq:set}). Here $k\simeq m\pi/a_r$ serves as the time, with $m=1,\cdots,n$. Similarly to the last subsection, the $k$th order Chebyshev polynomial $T_k(\mathcal{G})$ can be calculated. The cut-off order (evolution time) $K^{\prime}=\lfloor n\pi/a_r \rfloor$. More details can be found in Append.~\ref{app:2}. 

The duality of the Chebyshev polynomial, as being approximately the trigonometry functions and being the polynomials, plays a vital role in the DACP method. In a certain sense, the Chebyshev evolution is an efficient (possibly the most one among all the polynomials) simulation of the quantum oscillations~\cite{CHEN199919}.

\subsection{Subspace diagonalization}\label{subseq:dd}
Computing the basis $\left\{\left|\Psi_i\right>: i=1,\cdots,2n+1 \right\}$ (Eq.~(\ref{eq:set})) by combination of the exp-semicircle filter and the Chebyshev evolution represents the most challenging aspect as well as the most time-consuming part of the DACP method. Once the appropriate basis has been constructed, the task remained is straightforward, i.e., to compute the eigenpairs of the projected Hamiltonian $H$. This is equivalent to solving the generalized eigenvalue problem
\begin{equation}\label{eq:gep}
	HB=SB\Lambda.
\end{equation}
Here, $H$ and $S$ denote the projected Hamiltonian in $\mathbb{L}$ and the overlap matrices, respectively,
\begin{equation}
	H_{ij}=\left< \Psi_i\right| \mathcal{H}\left|\Psi_j\right>,
	S_{ij}=\left< \Psi_i| \Psi_j\right>.
\end{equation}
$\Lambda$ is a diagonal matrix with the eigenvalues in $[-a,a]$ and the matrix $B$ transforms the found basis Eq.~(\ref{eq:set}) to the eigenstates $\left|\phi_j\right>$ of $\mathcal{H}$,
\begin{equation}
	\left|\phi_j\right>=\sum_{i=1}^{2n+1}B_{ij}\left|\Psi_i\right>.
\end{equation}
All these matrices are of size $(2n+1)\times(2n+1)$. Because of the small size of it, $H$ can be explicitly expressed and stored in core memory. The necessary procedures are conveniently available in the LAPACK library~\cite{lapack}.

Importantly, due to the special property of the Chebyshev polynomial, the computation of matrices $H$ and $S$ can even be achieved without an explicit computation and storage of the states $\left|\Psi_i\right>$. This feature gives rise to a further improvement for the DACP method, both in computation time and memory saving. Besides, for an overcomplete basis Eq.~(\ref{eq:set}), the overlap matrix $S$ is generally singular. We thus employ the singular value decomposition (SVD) to solve it. After the SVD, the eigenvalues of $S$ are discarded if their absolute values sit below the cutoff condition $\varepsilon=10^{-12}$.  The number of remained eigenvalues effectively counts the linearly independent states in Eq.~(\ref{eq:set}).  We present both the explicit expressions (denoted by $T_k(\mathcal{G})$ and $\left|\psi_E \right>$ only) of matrices $H$ and $S$, and the solution of the generalized eigenvalue problem in Append.~\ref{app:3}.

Eigenvalues obtained by the subspace diagonalization may not own the same accuracy when compared to true eigenvalues of the system, thus it is necessary to conduct an independent check to estimate the error bounds. To this end, if the eigenstates $\left|\phi_j\right>$ were known, the residual norm (variance of the energy) $||r_j||=\sqrt{\left<\mathcal{H}^2\right>-\left<\mathcal{H}\right>^2}$, where $\left<\mathcal{H}^2\right>=\left<\phi_j\right|\mathcal{H}^2\left|\phi_j\right>$ and $\left<\mathcal{H}\right>=\left<\phi_j\right|\mathcal{H}\left|\phi_j\right>$, is widely used as the parameter measuring the accuracy of the results. It has been shown that $|| r_j||$ gives an upper bound on the true error (absolute error) of the computed eigenvalue~\cite{YSaad}.

\section{Numerical results} \label{sec:NR}

We apply the  DACP method to the quantum spin-$1/2$ systems with two-body interactions. Such systems are good models for investigating a large class of important problems in quantum computing, solid state theory, and quantum statistics~\cite{QCQI,IQSM,QPT,QPTT}. A large amount of exact eigenvalues helps us to obtain the statistical properties, to distinguish quantum chaos from integrability, and serves as a benchmark to evaluate other approximate methods as well.

Generally speaking, the DACP method can deal with the spin systems consisting of couplings between any pair of the $N$ spins. Each of the Pauli matrix $\sigma^{\alpha}$ or the two coupling Pauli matrices $\sigma^{\alpha}\otimes\sigma^{\beta}$, where $\alpha,\beta=x,y,z$, is properly represented by a specific function.

We specify the spin model for two physical systems. One is the disordered one-dimensional transverse field Ising model~\cite{PhysRevB.51.6411}, where the Hamiltonian is
\begin{equation}
{\mathcal{H}}=\frac{1}{4}\sum_{i=1}^{N-1}J_{i,i+1} \sigma_i^x \sigma_{i+1}^x+\frac{1}{2}\sum_{i=1}^N \Gamma_i^z \sigma_i^z,
\end{equation}
with $\sigma_i$ the Pauli matrices for the spin $i$. This system is exactly solvable by Jordan-Wigner transformation~\cite{QPT}, making it an ideal correctness checker for the DACP method. The nearest neighbor exchange interaction constants $J_{i,i+1}$ are random numbers that uniformly distributed in $[-J/\sqrt{N},J/\sqrt{N}]$ with $J=10$. The local random magnetic fields are represented by $\Gamma^z_i$, which are random numbers that uniformly distributed in the interval $[0,\Gamma]$ with $\Gamma=1$.

Another system is the spin glass shards~\cite{sgs}, which represents a class of global-range interacting systems that require relatively large bond dimensions to be tackled by the DMRG methods~\cite{RevModPhys.77.259}. The Hamiltonian describing the system is
\begin{equation}
{\mathcal{H}}=\sum_{ i<j}J_{ij}\sigma_i^x\sigma_{j}^x+\sum_i \Gamma_i^z\sigma_i^z.
\end{equation}
All symbols and parameters are the same as that of the above Ising model, except that the first summation runs over all possible spin pairs. This system is interesting because it presents two crossovers from integrability to quantum chaos and back to integrability again. In the limit $J / \Gamma \rightarrow 0$, the ground state is paramagnetic with all spins in the local field direction and the system is integrable~\cite{sgs}. In the opposite limit $J/\Gamma\rightarrow \infty$, the ground state is spin glass and the system is also integrable since there are $N$ operators ($\sigma_i^x$) commuting with the Hamiltonian. A quantum chaos region exists between these two limits. $J=10 \,\Gamma$ is approximately the border from the quantum chaos to the integrable (the spin glass side) when $N=20$~\cite{sgs}.

\begin{figure}[b]
	\includegraphics[width=3.25 in]{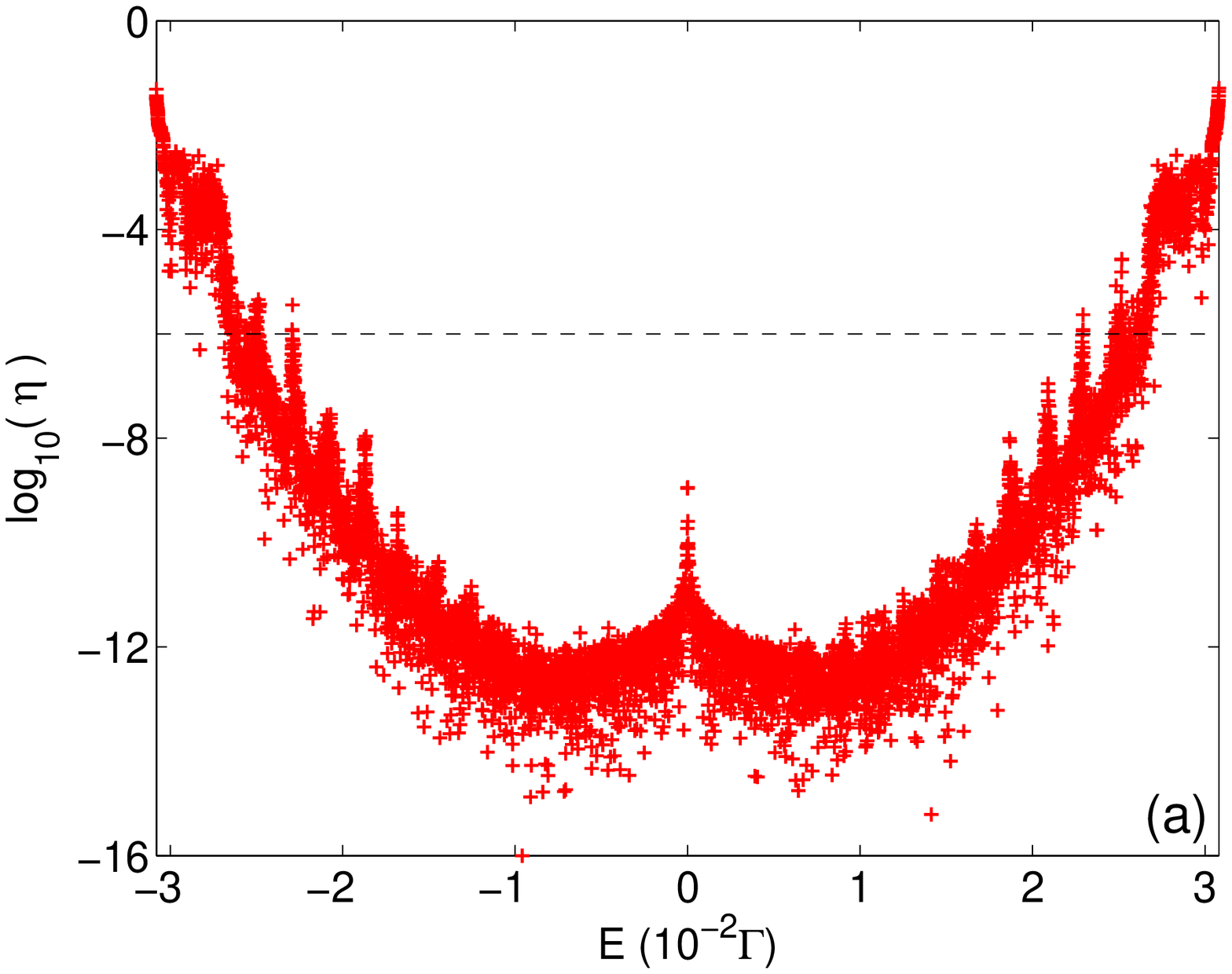}
	
	\includegraphics[width=3.25 in]{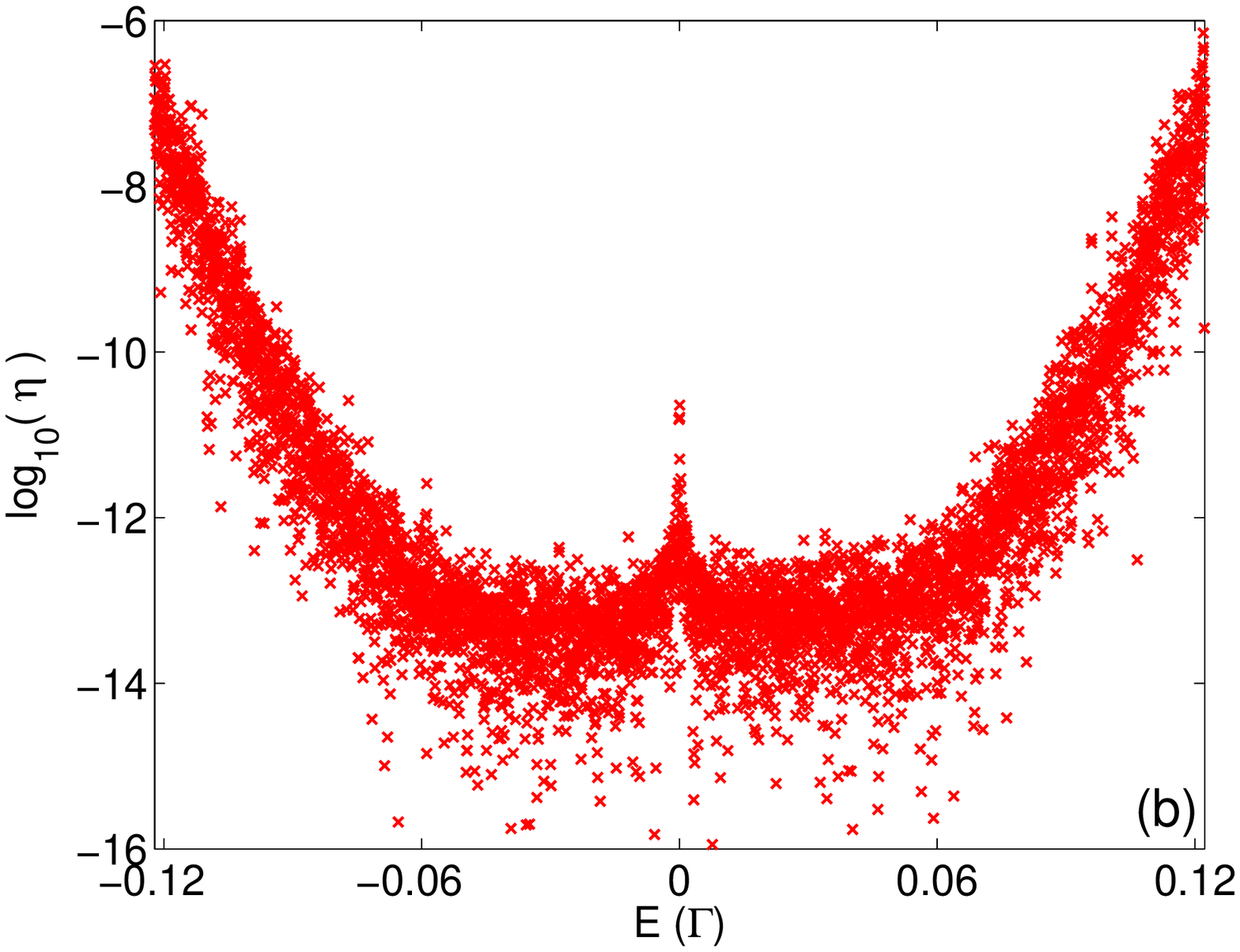}
	\caption{\label{fig:err} (Color online.) The relative error $\eta$ in logarithmic scale of the calculated eigenvalues, for (a) Ising model with $N=19$ and (b) spin glass shards model with $N=17$. The horizontal axis is the system energy. The number of eigenvalues satisfying $\eta<10^{-6}$ (black dashed line) is $5,385$ for (a) and $5,000$ for (b).}
\end{figure}

By employing the upper-bound-estimator, which costs little extra computation and bounds up the largest absolute eigenvalue $E_0$, one may estimate $E_{\rm max}=E_0$ and $E_{\rm min}=-E_0$~\cite{ZHOU2011480}. For this setting we have utilized
the symmetry of the density of states (DOS), a bell-shape profile centered at zero, in the many-spin systems. Since we require $5,000$ central eigenvalues, we may set $n=4,000$, corresponding to a dimension $8,001$ and being adequate to span the whole subspace $\mathbb{L}$. The overlap matrix $S$ is generally singular. The approximate distribution of DOS $\rho(E)$ may be efficiently calculated through the Fourier transformation of a time evolved wave function or through a better estimation method given in~\cite{dosestimation}. The parameter $a$ is  appropriately chosen to ensure that the number of eigenstates contained in $[-a,a]$ is a little less than $8,000$ (as illustrated in Fig.~\ref{fig:err}, the precision of some converged eigenvalues may be lower than required).

In practice, sometimes there may exist highly near-degenerate eigenvalues, with level spacings as small as $10^{-7}\Gamma$ while the average spacing is $10^{-5}\Gamma$. It is still hard (two magnitudes longer time) for the Chebyshev evolution to discriminate such close pair of eigenvalues. To circumvent this challenge, we employ the block filter technique~\cite{cd2}, which means a block of states is filtered or evolved ``simultaneously", in programming of the DACP method. The idea is that two or several random states in the degenerate subspace are usually linearly independent. For each numerical test, a block of $5$ initial trial states is randomly generated and employed with the parameter $n$ being adjusted to $n=800$ accordingly.

By these settings, we perform numerical tests on the above two systems to show the high exactness and efficiency of DACP method. For this work, we consider only the eigenvalues computations. All the timing information reported in this manuscript is obtained from calculations on the Intel(R) Xeon(R) CPU E7-8880 v4, using sequential mode.

In Fig.~\ref{fig:err} we present the relative error $\eta$ in logarithmic scale versus the system energy $E_{\rm exact}$, for the Ising model with $N=19$ and the spin glass shards with $N=17$. We have defined the relative error $\eta$ of the computed central eigenvalues $E$ as
$$\eta=\left| \frac{E-E_{\rm exact}}{E_{\rm exact}}\right|.$$
Exact eigenvalues of both systems have been obtained by other reliable methods. For the Ising model, we make use of the famous Jordan-Wigner transformation to reduce the original $2^N\times 2^N$ matrix to a $2N\times 2N$ one, and restore the full spectrum of the original Hamiltonian~\cite{QPT}. For the spin glass shards we simply utilize the function $eigs$ of MATLAB, to find $5,000$ eigenvalues closest to $E=0$. As for our numerical tests, the parameter $a$ in Fig.~\ref{fig:err} is $0.036\Gamma$ and $0.16\Gamma$ for (a) and (b), respectively.  In computing the Ising model, the number of eigenvalues satisfying $\eta<10^{-6}$ is not enough (less than $5,000$) by the settings mentioned above. By adjusting the block size to $10$ and the parameter $n$ to $500$, we then collect enough eigenvalues. The number of converged eigenvalues, i.e., computed eigenvalues satisfying the condition $\eta<10^{-6}$ is $5,385$ for (a) and $5,000$ (all the exact eigenvalues we have) for (b), while the total number of computed eigenvalues is $6,232$ and $5,910$, respectively.

The spike around $E=0$ for both figures is due to the smallness of the denominator $E_{\rm exact}$. The smallest absolute eigenvalue is about $4.4\times 10^{-6}\Gamma$ for (a) and $2.9\times 10^{-5}\Gamma$ for (b). Besides, there is a flat plateau in the middle of the figures, indicating that for those eigenvalues around $E=0$ we encounter the numerical error, i.e., the absolute error reaches the limit of the double precision representation. In Fig.~\ref{fig:err}(a) one may observe other spikes, which does not appear in (b). Each of those spikes corresponds to a cluster of eigenvalues lying extremely close to each other. The integrability of the Ising model implies the level clustering and a Poissonian distribution of energy level spacings. We speculate that the Chebyshev evolution does not function well for closely lied eigenvalues, as they give quite similar phase contributions to the final states. To significantly amplify the phase difference $\exp({\rm i}\Delta Et)$, it requires $t \gtrsim 1/\Delta E$, where $\Delta E$ is the energy difference. An alternative solution is to increase the block size to match up with the dimension of near-degenerate subspace. Ignoring the central plateau and the spikes, the distribution of $\eta$ shows an inverted shape of the exp-semicircle filter.

  \begin{figure}[b]
 	\includegraphics[width=3.25 in]{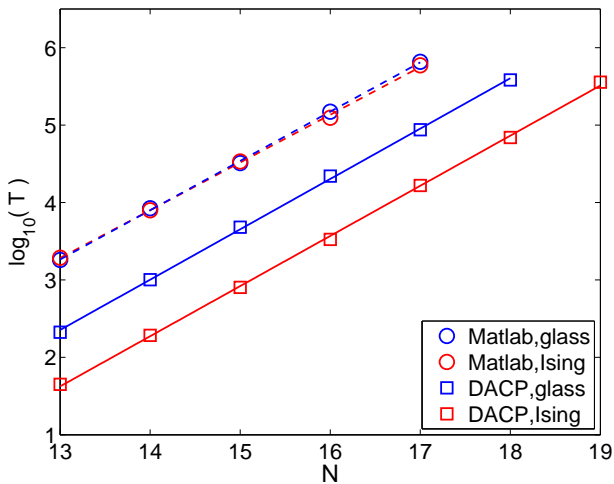}
 	\caption{\label{fig:scaling} (Color online.) Scaling behavior measured by the computation time $T$ (CPU seconds) in logarithmic scale, versus the system size $N$, with the DACP method (solid lines with squares) compared to the shift-invert method (dahed lines with circles) for the Ising model (red) and the spin glass shards (blue).  Each of the numerical tests, represented either by squares or circles, finds $5,000$ central eigenvalues. All the four lines are obtained via linear fitting. }
 \end{figure}

In Fig.~\ref{fig:scaling} we compare the computation time $T$ (CPU time in seconds) of the DACP method with that of the shift-invert approach. The latter is widely used in computing eigenpairs at the middle of the spectrum for quantum spin systems, to name a few, like in~\cite{scipost,PhysRevB.91.081103,PhysRevB.101.104201,PhysRevA.101.063617}. In our tests, it is implemented by the $eigs$ function of Matlab R2019b, which employs the implicitly restarted Arnoldi method (ARPACK)~\cite{Arpack}.
Each of the numerical tests finds around $5,000$ eigenvalues for either of the two spin systems. For the DACP method, we have discarded the time consumed during the subspace diagonalization, as it is a constant and does not affect the scaling behavior. Specifically, the subspace diagonalization spends roughly $700$ CPU seconds in each test. Such a constant is completely negligible for $N\ge 17$ systems.

As shown in Fig.~\ref{fig:scaling}, for spin systems with $13\le N \le 17$, the DACP method  is about $30$ times faster than the shift-invert approach for the Ising model and $8$ times faster for the spin glass shards. Since the shift-invert method essentially finds the ground energy of $\mathcal{H}^{-1}$, which is usually not a sparse matrix, its computation time $T$ is not affected by the sparsity of the Hamiltonian and the two dashed lines nearly coincide. On the contrary, the DACP method employs the polynomial combination of $\mathcal{H}$ acting on the states, its computation time $T$ is heavily influenced by the number of Pauli operators of the specific Hamiltonian. For example, when $N=18$ the computation time $T$ for the spin glass shards is $5.5$ times that for the Ising model, while the number of Pauli operators is $6.2$ times. Considering this effect, the DACP method is still advantageous over the shift-invert approach in the worst case where the  exchange interactions run over all possible spin pairs.

\begin{table}
	\caption{\label{tab:scaling} Scaling constants $\alpha$ by linear fitting of the four curves in Fig.~\ref{fig:scaling}.}
	\begin{tabular}{c|c|c}
		\hline\hline
		$\alpha$ & DACP method & Shift-invert method \\
		\hline
		Ising & 1.49 & 1.42  \\
		Glass & 1.50 & 1.47  \\
		\hline\hline
	\end{tabular}
\end{table}

Furthermore, the scaling among these two methods is comparable. To compare quantitatively, we extract the scaling constants $\alpha$ of the four lines by fitting the numerical results, where $\alpha$ is defined by $T=T_0\exp(\alpha N)$. The values of $\alpha$ are shown in Table~\ref{tab:scaling}. Indeed, the four scaling constants are quite close. Suppose that $\alpha$ remains unchanged for a bigger $N$, then the DACP method's advantage in time cost may keep on for rather large systems. In addition, as illustrated by the numerical tests of the shift-invert method in~\cite{scipost}, the factorization time for finding $\mathcal{H}^{-1}$ dominates other computation steps. Considering the factorization part only,  the execution time in~\cite{scipost} exhibits a scaling constant $\alpha\simeq 1.66$, indicating a worse scaling behavior for systems with large spins. The efficiency of the DACP method is thus confirmed.

  \begin{figure}[b]
	\includegraphics[width=3.25 in]{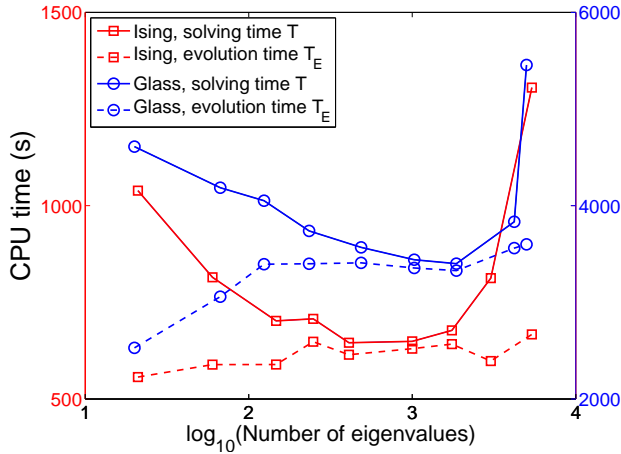}
	\caption{\label{fig:num} (Color online.) Computation time (CPU seconds) required versus number of  eigenvalues satisfying the condition $\eta<10^{-6}$, for the Ising model (red lines with squares, left axis) and the spin glass shards (blue lines with circles, right axis), using the DACP method. For both systems, $N=15$. The solid lines show the solving time $T=T_F+T_E+T_D$, including computation time for the exp-semicircle filtration $T_F$, Chebyshev evolution $T_E$ and subspace diagonalization $T_D$. The dashed lines present $T_E$.}
\end{figure}

In Fig.~\ref{fig:num} we show the computation time versus the number of converged eigenvalues, for the two systems of size $N=15$. The horizontal axis represents the number of computed eigenvalues satisfying the condition $\eta(E)< 10^{-6}$ in each test. Recall that the DACP method is divided into three parts: the exp-semicircle filtration, Chebyshev evolution, and subspace diagonalization. We denote their time consumption as $T_F$, $T_E$, and $T_D$, respectively, while solving time is the full computation time $T=T_F+T_E+T_D$. 
One immediately notice that for the two systems, computation time $T$ decreases on the left side, reaching a plateau at the middle region, while quickly increases on the right side. Yet the dashed lines $T_E$ present a quite small increasing rate for the whole region, as the horizontal axis is in logarithmic scale. In the middle region, where the DACP method performs the best,  the solving time $T$ and the evolution time $T_E$ roughly coincide with each other.

These features are easily understood if one considers the following four facts. (i) Notice that the DOS $\rho(E)$ is usually a bell-shape profile peaked at zero in spin systems and that $a$ is typically a tiny parameter ($a/E_{\rm max}\sim 10^{-2}$ for $N=19$), thus one may safely take $\rho(E)$ as a constant $\bar{\rho}$ in $[-a,a]$. Consequently, the number of required eigenvalues is $R \simeq 2\bar{\rho}a$. (ii) Setting the action of the Hamiltonian on the state, $\mathcal{H}\left|\psi\right>$, as a basic step, and denoting the computation time of the basic step as $\tau$, we may count the filtration time as $\tau$ times twice the cut-off order $K=18E_{\rm max}/a$, i.e., $T_F \simeq 36E_{\rm max}\tau / a \propto R^{-1}$. (iii) Similarly, we find the Chebyshev evolution time is $T_E \simeq \lfloor R\pi E_{\rm max}/a\rfloor \tau \propto R^0$ since $R\propto a$. (iv) It is well known that the full diagonalization time is proportional to the cube of the matrix size, i.e., $T_D \propto R^3$. The combination of these computation times clearly explains the behavior shown in Fig.~\ref{fig:num}. For the left side, the exp-semicircle filtration dominates, as $T_F$ is inversely proportional to the parameter $R$. Whereas, when it comes to the intermediate region where the number of eigenvalues is several hundreds to thousands, the evolution time $T_E$ consists of the majority of the computation time while the other two are negligible. As shown in Fig.~\ref{fig:num}, the performance of DACP method is approximately the same in finding $100$ to $3,000$ eigenvalues. As $R$ keeps increasing, the subspace diagonalization time $T_D$ eventually consumes the most computation time.

We remark that the plateau in the middle region in Fig.~\ref{fig:num} distinguishes the DACP method from those iterative filtering ones~\cite{PhysRevE.51.3643, Guan,Pieper2016High,cd1,cd2,dd}, although all of them take use of the Chebyshev polynomials. The iterative methods usually require a larger amount of filtration and reorthogonalization, thus a longer convergence time, in finding more eigenvalues~\cite{scipost,dd,PhysRevLett.125.156601}. In fact, as illustrated in Ref.~\cite{scipost}, one finds that the computation time is roughly proportional to the number of eigenvalues required for the shift-invert approach. The DACP method is thus highly desirable for large scale eigenvalues computations.

Finally, we note that the memory requirements by DACP method is rather small, as it works in a matrix-vector product mode which avoids an explicit matrix representation of the Hamiltonian. Moreover, as shown in Append.~\ref{app:3}, the whole set of states in Eq.~(\ref{eq:set}) is not preserved in memory. The major memory consumption is the storage of elements $H_{ij}$ and $S_{ij}$, thus the memory requirements of the DACP method relates only to the dimension $d$ of subspace $\mathbb{L}$. By the settings of this work, it occupies around $5.6$ GB of memory for appropriately calculating $5,000$ eigenvalues in $N\le 20$ systems. On the contrary, for the shift-invert method, the matrix size of $\mathcal{H}^{-1}$ grows rather fast as the system size $N$ increases (proportional to $4^N$), demanding a large amount of memory. For example, in our tests it consumes $573$ GB of memory for $N=17$ systems, which is already a factor of $100$ more than that of the DACP method.

\section{Conclusion} \label{sec:CON}

We propose the DACP method to efficiently calculate a large scale (at least $5,000$) of central eigenvalues with high precision, by employing twice the Chebyshev polynomials.  The proposed method is not a general eigensolver, as the generality gives way to the efficiency. It explores the excellent properties of Chebyshev polynomial to efficiently filter out those non-central parts of spectrum and to construct the appropriate basis states for the central spectrum. The numerical tests for the Ising model and the spin glass shards confirm the exactness and efficiency of the DACP method. Compared to the widely used shift-invert method, the DACP method gives a considerable increase in the speed of computations, for the Ising model up to a factor of $30$ while for the spin glass shards the increase in speed is less but still considerable (a factor of $8$). The memory requirement is drastically decreased,  up to a factor of $100$ for $N=17$ spin systems and even more for larger systems. Moreover, as shown in Append.~\ref{app:4}, the DACP method is both more stable and more efficient than the Lanczos method applied with the  exp-semicircle filter. As a powerful tool for central eigenvalues calculation, the DACP method may find potential applications in many physical problems, such as many-body localization in condensed matter physics~\cite{MBL,RevModPhys.91.021001,PhysRevB.91.081103,PhysRevA.101.063617}, and level statistics in quantum chaos~\cite{QuantumChaos, sgs,PhysRevB.101.104201,PhysRevE.102.020101}.

\appendix

\section{Chebyshev polynomial} \label{app:1}

In this paper, we limit ourselves to that only the polynomial combinations of $\mathcal{H}$ are the allowed operations. We utilize the Chebyshev polynomial to fulfill the tasks mentioned above. It is the key to bridge the combinations of $\mathcal{H}^k$ with an approximately exponential function of $\mathcal{H}$, either $\exp(k\mathcal{H})$ or $\exp({\rm i}k\mathcal{H})$. Due to its several remarkable properties, the Chebyshev polynomial may exhaust the potential of this type of methods.

\begin{figure}[b]
	\includegraphics[width=3.25 in]{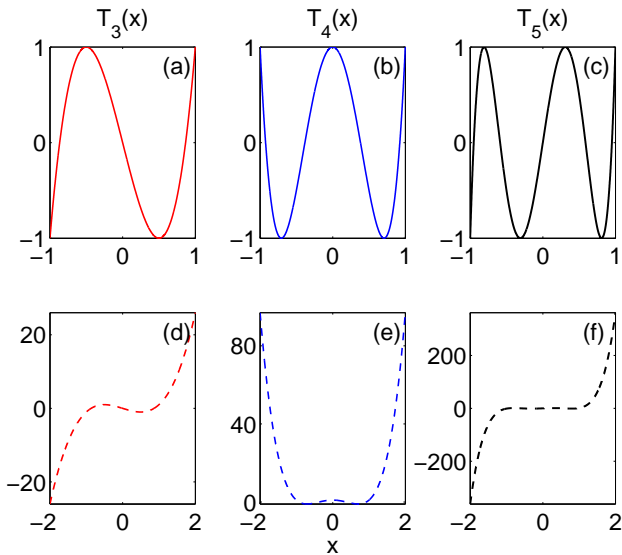}
	\caption{\label{fig:shape} (Color online.) Chebyshev polynomials $T_k(x)$ for $k=3$ (red lines), $k=4$ (blue lines), $k=5$ (black lines). The first row ((a-c) with solid lines) illustrate the oscillations of $T_k(x)$ inside the interval $[-1,1]$ while the second row ((d-f) with dashed lines)  show the rapid increase outside $[-1,1]$ of Chebyshev polynomials.}
\end{figure}

The $k$th order Chebyshev polynomial of the first kind is defined by
\begin{equation}
T_k\left( x \right) =\left\{ \begin{array}{l}
\cos\left( k \cos^{-1}\left( x \right) \right) , \ \ \ \ \ \ \ \ \ \ \ \ \ \ |x| \le 1\ \\
\cosh\left( k \cosh^{-1}\left( x \right) \right) ,\ \ \ \ \ \ \ \ \ \ \ \ x>1\ \ ,\\
\left( -1 \right) ^k\cosh\left( k \cosh^{-1}\left( -x \right)\right) ,\ x<-1 \\
\end{array} \right.
\label{eq:cheb}
\end{equation}
with initial conditions $T_0(x)=1$ and $T_1(x)=x$~\cite{Handscomb2003Chebyshev}. It is a piece-wise function containing two different kinds of expression, while being the polynomial function, it is both continuous and smooth. For simplicity, let us set $\theta=\cos^{-1}(x)$ ($\cos \theta=x$) when $x\in [-1,1]$ and set $\theta=\cosh^{-1}(x)$ when $x\in [1,\infty)$, the corresponding range of $\theta$ is $\theta\in [0,\pi]$ and $\theta\in [0,\infty)$, respectively. In terms of the variable $\theta$, Eq.~(\ref{eq:cheb}) becomes
\begin{equation}
T_k\left( x \right) =\left\{ \begin{array}{l}
\cos\left( k \theta\right) , \ \ \ \ \ \ \ \ \ \ \ \ \ |x| \le 1\ \\
\cosh\left( k \theta\right) ,\ \ \ \ \ \ \ \ \ \ \ \ \ x>1.\ \\
\end{array} \right.
\label{eq:cheb2}
\end{equation}
One may easily observe that $T_k(x)$ is a sine or cosine-like oscillation function bounded by $-1$ and $1$ inside the interval $[-1,1]$, as illustrated in Fig.~\ref{fig:shape} (a-c), while it grows extremely fast outside $[-1,1]$, as shown in Fig.~\ref{fig:shape} (d-f).

Note that $\cosh(k\theta)=({\rm e}^{k\theta}+{\rm e}^{-k\theta})/2$. It is natural to expect an exponential growth of the Chebyshev polynomial outside the interval $[-1,1]$. In fact, it is known that among all polynomials with degree $\le k$, the Chebyshev polynomial $T_k(x)$ grows the fastest outside the interval $[-1,1]$ under comparable conditions~\cite{Rivlin}.

Associated with those properties is a practically useful one: $T_{k+1}(x)$ can be efficiently determined by using the $3$-term recurrence
\begin{equation}
T_{k+1}(x)=2xT_k(x)-T_{k-1}(x).
\label{eq:rec}
\end{equation}
All these properties of the Chebyshev polynomial render it a powerful toolbox and are  of great use for the DACP method.

\section{Detailed deductions of Eq.~(\ref{eq:set})}\label{app:2}

Here we derive explicit expressions showing how to construct the set Eq.~(\ref{eq:set}) via the Chebyshev evolution.
 We focus on the case that $a\ll E_{\rm max}$, which is fairly reasonable for large ($N\ge 15$) spin systems.
\begin{equation}\label{eq:Acheb}
\begin{aligned}
\left| \psi_E \left( k \right) \right>
&=T_k(\mathcal{G})\left| \psi_E  \right>\\
&=\sum_{j}\cos(k\omega_j)c_j^{\prime}\left| \phi_j\right>\\
&\simeq \sum_{j}\cos(\frac{k\pi}{2}-\frac{kE_j}{E_{\rm max}})c_j^{\prime}\left| \phi_j\right>\\
&=\left\{
\begin{array}{l}
(-1)^n\sum_{j}\cos(k\frac{E_j}{E_{\rm max}})c_j^{\prime}\left| \phi_j\right>, \quad  k=2n\\
\\
(-1)^n\sum_{j}\sin(k\frac{E_j}{E_{\rm max}})c_j^{\prime}\left| \phi_j\right>, \quad  k=2n+1.\\
\end{array}\right.
\end{aligned}
\end{equation}
Using the fact that when $x$ is small, $\arccos(x)=\pi/2-x+o(x)$, we have $\omega_j=\arccos(E_j/E_{\rm max})\simeq \pi/2-E_j/E_{\rm max}$ at the third line of Eq.~(\ref{eq:Acheb}). Therefore, we have the expression
 \begin{equation}\label{eq:chebevo}
 	T_k(\mathcal{G})\simeq\left\{
 	\begin{array}{l}
 	(-1)^n\cos(k\mathcal{G}), \quad  k=2n\\
 	(-1)^n\sin(k\mathcal{G}), \quad  k=2n+1.
 	\end{array}\right.
 \end{equation}
 We then conduct a Chebyshev evolution with a cutoff order $K^{\prime}=\lfloor n\pi/a_r \rfloor$, recording both $T_{k-1}(\mathcal{G})\left| \psi_E \right>$ and $T_{k}(\mathcal{G})\left| \psi_E \right>$ when $k=\lfloor m\pi/a_r \rfloor$ with $m=1,\cdots,n$. After the evolution is done, the set of states Eq.~(\ref{eq:set}) is automatically generated.

\section{Evaluation of matrix elements and solution of the generalized eigenvalue problem }\label{app:3}

As shown in Append.~\ref{app:2}, we may rewrite the basis Eq.~(\ref{eq:set}), or $\left\{\left|\Psi_i\right>: i=1,\cdots,2n+1 \right\}$, using the Chebyshev polynomials:
 \begin{equation}\label{eq:set2}
 \left\{ I,T_{k_1-1}(\mathcal{G}),T_{k_1}(\mathcal{G})\cdots,T_{k_n-1}(\mathcal{G}),T_{k_n}(\mathcal{G}) \right\} \left| \psi_E \right>,
 \end{equation}
 where $k_m=\lfloor m\pi/a_r \rfloor, \ m=1,\cdots, n$, $\ket{\Psi_1} =\ket{\psi_E}$,
 $\left|\Psi_2\right> = T_{k_1-1}(\mathcal{G})\left|\psi_E\right>$, 
 $\left|\Psi_3\right> = T_{k_1}(\mathcal{G})\left|\psi_E\right>$, etc. For simplicity, one may further assume $k_m$ is an even number.

The elements $S_{ij}=\left< \Psi_i |\Psi_j\right>=\left< \psi_E\right|T_x(\mathcal{G}) T_y(\mathcal{G})\left|\psi_E\right>$, where $x$ and $y$ are directly determined by $i$ and $j$. The correspondence is one to one. By making use of the relation
\begin{equation}
T_i(\mathcal{H})T_j(\mathcal{H})=\frac{1}{2}(T_{i+j}(\mathcal{H})+T_{|i-j|}(\mathcal{H})),
\label{eq:shift}
\end{equation}
one may even find the matrix elements without recording any states during the Chebyshev evolution.  Instead, we simply record the two numbers $\left< \psi_E\right|T_k(\mathcal{G})\left|\psi_E\right>$ and $\left< \psi_E\right|\mathcal{H}T_k(\mathcal{G})\left|\psi_E\right>$ at the appropriate time, i.e., when
$k=k_m-2$, $k=k_m-1$, $k=k_m$ and $k=k_m+1$.

Finally, we arrive at the explicit expressions of matrix elements
\begin{gather}
	S_{ij}=\frac{1}{2} \left< \psi_E\right|\left(T_{x+y}(\mathcal{G})+T_{|x-y|}(\mathcal{G})\right)\left|\psi_E\right>, 	\\
H_{ij}=\frac{1}{2} \left< \psi_E\right|\mathcal{H}\left(T_{x+y}(\mathcal{G})+T_{|x-y|}(\mathcal{G})\right)\left|\psi_E\right>.
\end{gather}
Also note that since $T_{x+y}$ is needed, where both $x$ and $y$ may reach the maximum value $k_n$, the cut-off order $K$ is doubled to $K=2k_n$ in this mode.

For the generalized eigenvalue problem Eq.~(\ref{eq:gep}), the Hermitian matrix $S$ is first diagonalized as
\begin{equation}
	S=V\Lambda _sV^{\dagger},
\end{equation}
where $V$ is the eigenvector matrix for $S$, $VV^{\dagger}=I$, and $\Lambda_s$ is the associated eigenvalue matrix. Since $S$ is generally singular, we then contract the $(2n+1)\times(2n+1)$ matrix $V$ by elimination of the columns associated with eigenvalues with absolute value below a cutoff $\varepsilon=10^{-12}$. Denoting the number of retained eigenvalues as $m$, the contracted eigenvector matrix $\widetilde{V}$ is of order $(2n+1)\times m$, and
\begin{equation}
	\widetilde{S}=\widetilde{V}\widetilde{\Lambda} _s\widetilde{V}^{\dagger}.
\end{equation}
 The next step is to form the contracted Hamiltonian matrix $\widetilde{H}$. Since
 \begin{equation}
 	I=\left(\widetilde{\Lambda} _s^{-\frac12}\widetilde{V}^{\dagger}\right) \widetilde{S} \left(\widetilde{V}\widetilde{\Lambda} _s^{-\frac12}\right),
 \end{equation}
denoting the transformation matrix $U=\widetilde{V}\widetilde{\Lambda} _s^{-\frac12}$, the contracted $m\times m$ Hamiltonian matrix is
\begin{equation}
	\widetilde{{H}}=U^{\dagger}HU.
\end{equation}
The Hermitian matrix $\widetilde{H}$ of order $m\times m$ with $m$ around $10^3$ to $10^4$ is then diagonalized directly
\begin{equation}
	\widetilde{H}=\widetilde{Y}\widetilde{\Lambda}\widetilde{Y}^{\dagger} ,
\end{equation}
where $\widetilde{Y}$ is the eigenvector matrix of $\widetilde{H}$ and $\widetilde{\Lambda}$ is a diagonal matrix consisting of those desired eigenvalues of the original Hamiltonian $\mathcal{H}$ contained in $[-a,a]$. The eigenstates of the projected Hamiltonian $H$ may be obtained through elementary matrix algebra:
\begin{equation}
	B=U\widetilde{Y}=\widetilde{V}\widetilde{\Lambda} _s^{-\frac12}\widetilde{Y}.
\end{equation}
Denoting the Eq.~(\ref{eq:set2}) as a $2^N\times(2n+1)$ matrix $A$ with $\left|\chi_i\right>$ being the $i$-th column, the eigenstates of the original Hamiltonian $\mathcal{H}$ contained in $[-a,a]$ is
\begin{equation}
	\Phi=AB=A\widetilde{V}\widetilde{\Lambda} _s^{-\frac12}\widetilde{Y}.
\end{equation}
We finally get the answers $\widetilde{\Lambda}$ and $\Phi$.

\section{Comparison with the Lanczos method}\label{app:4}
Since the Lanczos method is widely employed to compute the lowest (or highest) eigenvalues, while the exp-semicircle filter provides a means to transform the central eigenvalues to the highest ones, naturally one may combine them together to efficiently compute the central eigenvalues, a way far simpler than the DACP method. But there are several reasons to prefer complicated techniques used in the DACP method to the standard construction of the Krylov space, especially when large scale computations are required.

First, it is known that the Lanczos method is not numerically stable under practical conditions. Specifically, although the Lanczos algorithm shows perfect properties in theory, in practice it loses many of its designed features, e.g., global orthogonality
and linear independence among Lanczos recursion states~\cite{2003On,1985Lanczos,YSaad}. These defects effectively limits the number of eigenvalues which can be computed. In addition, it seems that the emergence of generalized eigenvalue problem (to deal with non-orthogonal base states) is unavoidable due to the error accumulations~\cite{book:1247092, doi:10.1063/1.468999}. The Chebyshev recursion, on the other hand, possesses many interesting properties common in both the ideal and practical calculations~\cite{2003On}. In particular, it is accurate and stable for $x\in[-1,1]$, allowing the propagation in the Chebyshev order domain for tens of thousands of steps without accumulating significant errors~\cite{CHEN199919}.

Second, taking the reorthogonalization step into consideration, the efficiency of the Lanczos algorithm decreases. The reorthogonalization step is a necessary part in both the Lanczos and the Arnoldi method. When there is a large amount (e.g., several thousands) of eigenvalues to be computed, the total cost is actually dominated by the reorthogonalization~\cite{cd1,cd2,2010Diagonalization}. Suppose the total required number of eigenvalues is $R$ and the Hilbert space dimension is $D$, then the Lanczos-type methods scale as $DR^2$, as every generated Ritz vector needs to be reorthogonalized against the existing ones (see also the discussions in Refs.~\cite{cd2,Arpack}). In comparison, we have shown in Sec.~\ref{sec:NR} that $T_F\propto DR^{-1}$, $T_E\propto DR^0$, and $T_D\propto D^0R^3$, where each one of them is better than $DR^2$ ($R\ll D$). Although partial reorthogonalization schemes have been proposed, they result in an increased cost in computations as well as memory traffic~\cite{dd,doi:10.1137/060675435}, and they are not guaranteed to succeed when the accuracy requirement becomes strict~\cite{doi:10.1137/060675435}. Moreover, the DACP method and the Lanczos algorithm are different in terms of space complexity, as the former shows $\max(DR^0,R^2)$, while the latter requires $DR$ once the reorthogonalizations are needed~\cite{Arpack}. Therefore, the DACP method has superiority over the Lanczos algorithm in both the time and the space complexity.

Finally, ignoring the requirement of reorthogonalizations, we find that the two methods are comparable in both the time and the space complexity. Recall that the Chebyshev evolution time, being the dominate term in the DACP method, becomes $T_E\simeq\pi R E_{max}\tau/a$, where $\tau$ is the time for matrix-vector product. We then discuss the time complexity of a combined version, the Lanczos method with the exp-semicircle filter. As shown in Fig. 1 of the Supplemental Material in Ref.~\cite{PhysRevLett.125.156601}, and the discussions in Ref.~\cite{cd1}, the number of Lanczos steps $m$ shall be at least twice the number of requested eigenvalues $R$, i.e., $m\ge 2R$. Since $2R$ iterations are needed, and every iteration requires a filter application with $T_F\simeq 36E_{max}\tau/a$, the computation time reads $2R\cdot 36E_{max}\tau/a =72RE_{max}\tau/a$. Therefore, the DACP method and the combination of Lanczos with the exp-semicircle filter share a same time complexity.

In summary, the DACP method is both more stable and more efficient than the Lanczos method applied with the exp-semicircle filter.
\addcontentsline{toc}{chapter}{Acknowledgment}

\section*{Acknowledgment}

We thank X.-H. Deng for discussions. This work is supported by the NSFC Grant No. 91836101, No. U1930201, and No. 11574239. The numerical calculations in this paper have been done on the supercomputing system in the Supercomputing Center of Wuhan University.


\providecommand{\noopsort}[1]{}\providecommand{\singleletter}[1]{#1}%

\end{document}